# Methods for quantum interference in atomic ensemble

Niladri Ghorui[1], Sudip Mandal[1] and Swarupananda Pradhan[1,2,3]

[1] Photonics and Quantum optics section, Atomic and molecular physics division, Bhabha Atomic Research Centre Facilities - Visakhapatnam-531011, India

[2] Homi Bhabha National Institute, Department of Atomic Energy- Mumbai-400094, India

[3] Corresponding author: spradhan@barc.gov.in

**Abstract**

An experimental method for obtaining quantum interference signal in atomic ensemble using a bi-chromatic field is discussed. Here, the quantum interference signal is obtained by scanning the magnetic field rather than conventional method of changing the frequency separation between the light fields. We could simultaneously observe resonances due to population redistribution and quantum superposition between non-degenerate states, which otherwise involves fundamentally different approaches. The method is implemented to Rubidium atoms in buffer gas filled as well as anti-relaxation coated atomic cells. Apart from phenomenological interest, the modified experimental procedure is found to be convenient for *in-situ* calibration of three axis magnetic coils. The investigation will be useful for high as well as low vector magnetic field sensing.

## I. Introduction:

The developments of various kinds of quantum devices are integral part of the on-going second quantum revolution.[1] These quantum sensors and standards exhibit optimal performance under specific ambient conditions.[2,3] For example, standard like atomic clock operates with a C-field to reduce the influence of magnetic field.[4] The coherent population trapping (CPT) based atomic magnetometer requires a bias magnetic field to measure shift in the magnetic resonance.[5] The Hanle based atomic magnetometers operates near zero magnetic field and thus, proper cancelation of the background magnetic field is very important.[6] The required field environment for the operation of the device is created by three axis magnetic coils. The ambient magnetic field components are measured from the bias currents flowing through the coils and constitute basic part of magnetic sensing. Thus, experimental methods for accurate calibration of the magnetic coils are central to these applications.

One of the advantages of the atomic magnetometer/ devices is *in-situ* calibration of the magnetic coils. It avoids the need of a reference magnetometer and accuracy is limited by the knowledge of fixed atomic constants. In this article, we have explored a modified experimental procedure for acquiring quantum interference signal in atomic ensemble that is found to be convenient for calibration of the magnetic coils. The quantum interference in atom comes in to play as an (common) atomic level gets coupled to different level (competing pathways) through light fields. It gives rise to CPT phenomena with unusual characteristic. In conventional experimental configuration, the frequency separation of a bi-chromatic field is scanned by changing the frequency of a radio frequency (RF) oscillator while keeping the atomic ensemble at a fixed bias magnetic field. The acquired resonances are used in atomic magnetometer, atomic clock, and other sensors.[3,5,7] In this article, we have explored a contrasting approach where the quantum interference signal is acquired by scanning the magnetic field while keeping the frequency separation between the light fields (frequency of RF oscillator) fixed. It leads to topsy-turvy in the position of the resonances. The modified method offers several advantages over the conventional technique.

A comparison on the characteristic resonances acquired using the conventional and the modified method, both utilizing a bi-chromatic light field is presented. The attributes of the resonances obtained using the modified method in buffer-gas filled (BF) as well as anti-relaxation coated (ARC) atomic cell, along with complementary signal extraction methods are discussed. The calibration of the magnetic coils is found to be simplified in the modified method. More importantly, resonances involving quantum interference between non-degenerate states and population redistribution/quantum interference among degenerate states are simultaneously acquired. It is in contrast to erstwhile involvement of two different experimental approaches.[5,6]

## II. Experimental details:

The experiments are carried out on Rb atoms with natural isotopic composition in glass cells. The BF cell (50 mm length) contains atomic sample at 25 torr $N_2$ buffer gas.[5-8] Whereas, octadecyltrichlorosilane is coated on the inner surface of ARC cell with 25 mm length for realizing anti-relaxation of atomic states.[9-11] The schematic of the experimental set up is shown in Fig.-1.

A vertical cavity surface emitting diode laser (VCSEL) with ~60 µW power and ~40 MHz line-width is utilized. The $^{85}Rb$ atoms are coupled to the light field with $D_1$ transition at 795nm. The VCSEL frequency is modulated at ~1.517 GHz to generate side modes for establishing superposition of states originating from separate ground hyperfine levels. Around 10% of the laser beam is tapped by a beam splitter (BS) for frequency stabilization of the laser using Rb spectrometer. A frequency modulation (FM) spectroscopy set-up comprising of Rb atomic cell, local oscillator (12 kHz), photo detector and lock-in amplifier (LIA) represents the Rb spectrometer. The laser is frequency modulated by the local oscillator and the photodiode signal in the Rb spectrometer is demodulated for generating FM signal for frequency stabilization of the laser. The power of the main beam is controlled by a half wave plate (HWP) and a polarization beam splitter cube (PBS). The ambient magnetic field around the atomic cell is shielded by four layer of mu-metal shield. The remnant (bias) magnetic field along three orthogonal directions are compensated (applied) by controlling the currents through three set of Helmholtz coil.

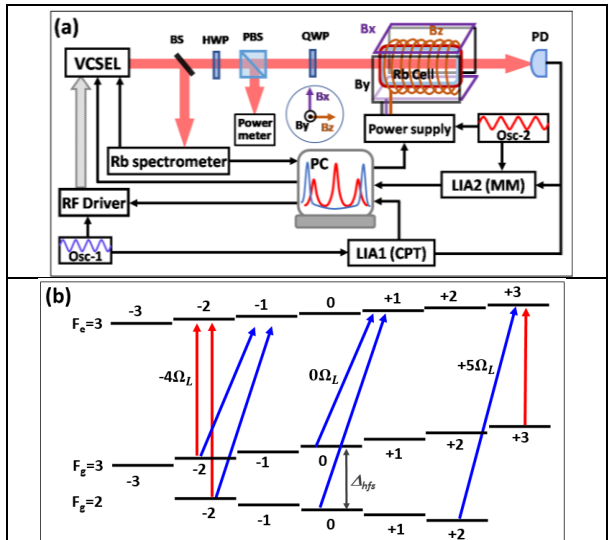

Fig.-1(a): Schematics of the experimental set-up. The VCSEL is frequency modulated by a RF driver (~1.517 GHz). The RF itself is frequency modulated at 440 Hz (Osc-1). The magnetic fields (one at a time) are modulated at 39 Hz (Osc-2). The PD signal is phase-sensitively detected with lock-in amplifiers (LIA-1 and LIA-2). The resultant signals are termed CPT and $MM$ signal. (b): The schematic of coupling between atomic states with the light fields. Both σ and π transition are allowed owing to the presence of transverse magnetic field. Some of the couplings giving rise to $-4\Omega_L$, $0\Omega_L$ (clock transition) and $+5\Omega_L$ (ideal for magnetometry) are shown.

The RF is modulated at 440 Hz using a local oscillator (Osc-1). The photo detector (PD) signal, filtered with LIA-1 in reference to Osc-1 is termed as CPT signal. The magnetic field is modulated with Osc-2 (39 Hz). The PD signal demodulated with LIA-2 with reference to Osc-2 is named $MM$ signal. The subscripts $x$, $y$ and $z$ are used in $MM$ to denote the axis of magnetic field modulation. The quarter-wave plate (QWP) before the Rb cell polarizes (right circularly $\sigma^+$) the light field. The data acquisition, laser frequency stabilization, RF control, and magnetic field scanning are carried out by a personal computer (PC).

### III. Results and discussion:

The Larmor's frequency ($\Omega_L$) of the atomic spin precision is given by

$$\Omega_L = g_F \mu_B B \text{-----------------------------(1)}$$

Where $g_F$ is the gyromagnetic factor, $\mu_B$ is the Bohr magneton, total magnetic field $B = \sqrt{B_x^2 + B_y^2 + B_z^2}$, $B_x$ and $B_y$ are magnetic field components transverse to the light propagation direction and $B_z$ is the longitudinal magnetic field. The light polarization is kept right circularly polarized, hence only $\sigma^+$ couplings are allowed in absence of transverse magnetic field. However in presence of a transverse magnetic field, $\pi$ couplings are allowed in addition to $\sigma$ couplings. Some of the couplings are shown in Fig.-1(b). The $\sigma^-$ coupling arising due to transverse magnetic field are feeble and not shown. Consequently, CPT resonances at both even and odd harmonics of $\Omega_L$ are observed in presence of transverse magnetic field as shown in Fig. 2(a). Here, the CPT signal is obtained by scanning RF frequency and PD signal is phase sensitively detected in reference to the modulation (Osc-1) applied to the RF. A resonance appears when the frequency separation between the two light fields (Fig-1(b)) becomes equal to the separation between the ground Zeeman states. The two photon resonance condition can be written as

$$2 \times \nu_{RF} = \Delta_{hfs} \pm n \times \Omega_L \text{---------------------(2)}$$

Where $\nu_{RF}$ is the frequency of the RF oscillator. The factor 2 arises as the separation between the utilized light fields is $2 \times \nu_{RF}$ due to the participation of $\pm 1$ side modes in the quantum interference phenomenon. We define RF detuning ($\delta_{RF}$) as

$$\delta_{RF} = 2 \times \nu_{RF} - \Delta_{hfs} \text{------------------------(3)}$$

The possible values of $n$ ranges from -5 to +5 for $^{85}Rb$ atoms. The frequency separation between the light fields is controlled by the frequency of the RF. One of the prominent advantages on the CPT based devices is that the magnetic coils (used in the experiment) can be *in situ* calibrated using Eq. 1&2. The $\Omega_L$ being linear in $B$ with fixed atomic parameters ($g_F, \mu_B$) as the proportionality constant, its measurement provides an accurate method for calibration of the magnetic coils. The $\Omega_L$ is measured from the RF frequency separation between two adjacent resonances in Fig.-2(a). However, the measurement of field components is still an involved task as the separation between the resonances depends on the total magnetic field $B$.

Fig-2: (a) *Conventional method:* The experimentally observed CPT resonances as a function of RF frequency. The experiment is done in presence of both longitudinal and transverse magnetic field. Hence, resonances at even as well as odd harmonics of $\Omega_L$ are observed. (b) *Modified method with magnetic demodulation:* The *MM* signal as a function of $B_z$ with $\sqrt{B_x^2 + B_y^2}$~+13 µT (blue curve) and ~0 µT (black curve) while keeping the RF detuning ($\delta_{RF}$) ~-434 kHz. The amplitude of blue curve is 5 times magnified with respect to black curve. A zoomed portion of the black curve is in the insert (c) *Modified method with RF demodulation:* The CPT signal as a function of $B_z$ with $\sqrt{B_x^2 + B_y^2}$~+13µT and $\delta_{RF}$ ~-434 kHz. The position of the resonances is marked using Eq.-2.

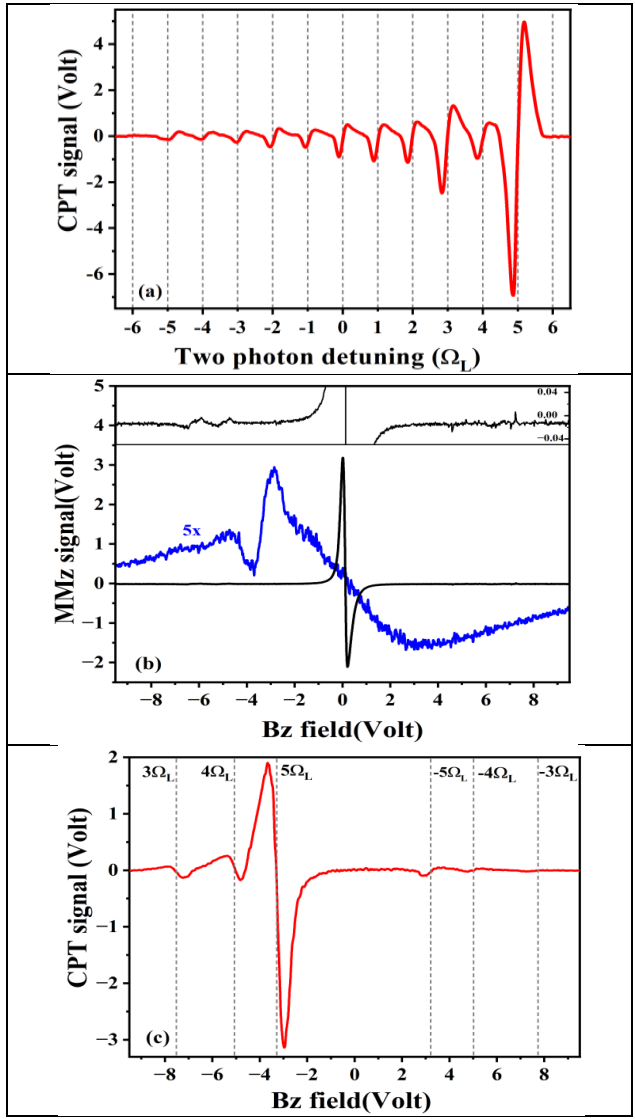

The Fig.-2 compares (a) resonances obtained through conventional method of scanning $\delta_{RF}$ at fixed $B$ with (b) & (c) resonances acquired by the modified method of scanning $B$ at fixed $\delta_{RF}$. Fig.-2(b) shows the *MM* signal as a function of $B_z$ field for $\sqrt{B_x^2 + B_y^2}$ zero and non-zero values. The $\delta_{RF}$ is kept at ~-434 kHz. For $\sqrt{B_x^2 + B_y^2}$ ~0 µT, a single dispersive profile corresponding to enhanced absorption (black curve in Fig. 2(b)) is observed. It arises due to population redistribution at zero magnetic field, similar to the observation involving monochromatic light.[12-14] However, both ground levels are coupled to the excited level against a single ground level for the erstwhile experiments involving monochromatic light. The two photon resonances (following Eq.-2) have feeble amplitude and can be observed in the zoomed plot (insert). The +5 $\Omega_L$ resonance gets prominent (blue curve in Fig.-2(b)) on increasing the $\sqrt{B_x^2 + B_y^2}$ to ~+13 µT. Similar enhancement of +5 $\Omega_L$ has been reported in recent literature, though using conventional method of RF scanning.[15] The modified method allows simultaneous acquisition of resonances due to zero field population redistribution among degenerate states and quantum interference signal between non-degenerate states. The broad population redistribution signal is due to increase in the amplitude of transverse magnetic field that also leads to decrease in its amplitude.

The +5 $\Omega_L$ has highest sensitivity to the magnetic field and is ideal for magnetometry. It has optimum amplitude at bias field of $B_z = \sqrt{B_x^2 + B_y^2}$ as has been pointed out in past literature using conventional method.[15] However in the modified method, the population redistribution signal overlaps with the +5 $\Omega_L$ resonance, limiting its prospect for high sensitive magnetometry. The overlapping is circumvented by the CPT signal, obtained by demodulating the PD signal with respect to the modulation applied to $\nu_{RF}$. Since the population redistribution signal is not dependent on $\nu_{RF}$, it doesn't contribute to the demodulated signal. The solo quantum interference resonances (following Eq.-2) are shown in Fig.-2 (c) following the modified method. It is interesting to note the change in the position of the resonances as compared to Fig.-2(a). The $\pm 1\Omega_L$ resonances will be at $\mp 91$ µT and are beyond the scanning range of our existing set-up. The clock transition (central $0\Omega_L$ resonance in Fig.-2(a)) being magnetically insensitive, cannot be observed by this method of scanning $B$ field.

One of the advantages of the modified method is estimation of the background magnetic field. The symmetry of $\pm n\Omega_L$ resonances in Fig.-2(c) with respect to the zero magnetic field along the scanning direction is not affected by the presence of magnetic field in other direction. It provides effortless estimation of the background magnetic field in any direction by simply scanning the magnetic field along it.

The procedure for calibrating the magnetic coil starts with compensation of background magnetic field in all directions. The CPT signals are acquired by scanning the current passing through the coil (to be calibrated). The CPT resonances and $MM$ signal (after compensation of the background $B$ field) are compared in Fig.-3. Here, the $\delta_{RF}$ is fixed at -58 kHz. The Lambda (Λ) system for $\sigma^+$ light arises from ground states having same value of $m_F$. Thus, four CPT resonances are observed for scanning of the $B_z$ field as only $\sigma^+$ coupling are allowed in absence of transverse magnetic fields.[16,17] These resonances appear at even multiples of $\Omega_L$ i.e. $\pm 4\Omega_L$ (~∓3µT) and $\pm 2\Omega_L$ (~∓6 µT). For the scanning of $B_x$ or $B_y$ field, $\pi$ couplings are also allowed and the CPT resonances at odd harmonics of $\Omega_L$ i.e. $\pm 3\Omega_L$ and $\pm 5\Omega_L$ are observed.[15,18-20] The $\pm 1\Omega_L$ resonances will be at $\mp 12$µT and are beyond the scanning range of our existing set-up. The presence of $\pm 1\Omega_L$ resonances is verified (not shown here) by lowering the value of $|\delta_{RF}|$. The even resonances have feeble amplitude and are not observed. The CPT signal in Fig.-3 allows calibration of the magnetic coils using Eq. 2 & 3. In contrast to this simple procedure, the estimation of the background field and calibration of the magnetic coil are an involved process in the conventional method where signal at multiple bias magnetic field are required.

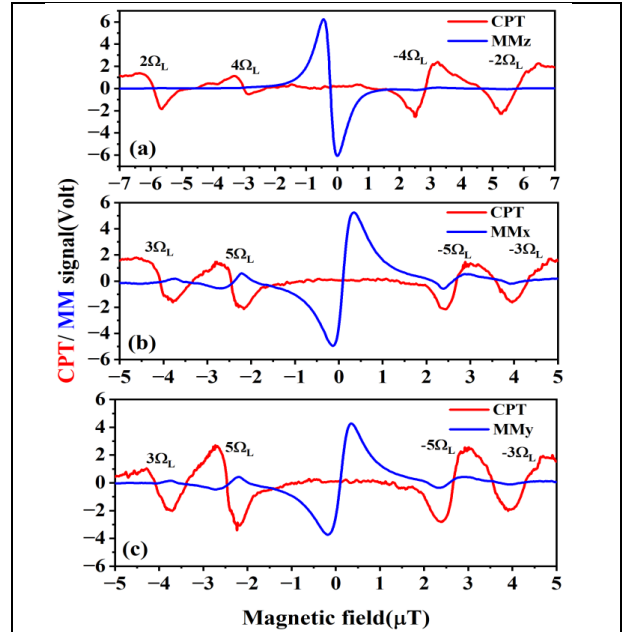

Fig-3: The experimentally observed CPT resonances in BF cell for scanning of $B_x$, $B_y$, and $B_z$ field. The fields orthogonal to the scanning direction are kept at zero. The $B_z$ scanning shows resonances at even multiples of $\Omega_L$ (a), whereas only odd multiples of $\Omega_L$ are observed for $B_x$ (b) and $B_y$ (c) scanning. The RF detuning ($\delta_{RF}$) is kept fixed at -58kHz. The sensitivity of lock-in amplifiers are $300\mu V$ and $1mV$ for CPT and $MM$ signals respectively.

The ARC cell provides a complementary platform for increasing the lifetime of coherently prepared superposition of states without incurring large inhomogeneous broadening encountered in buffer gas filled cell.[21] The modified method of scanning of magnetic field at fixed $\delta_{RF}$(~-316 kHz) shows clean quantum interference signal along with population redistribution resonance in the $MM$ signal as shown in Fig.-4(a). It may be compared with the black curve for buffer gas filled cell in Fig.-2(b), 3(a) where the quantum interference resonances have extremely feeble presence. The prominent observation of the quantum interference resonance in ARC cell is attributed to the diminishing collision compared to BF cell. In contrast, the population redistribution resonance is more favourable in BF cell due to longer confinement of the atoms in the laser beam.[21] The absorptive population redistribution resonance in ARC cell changes to transmission (changes polarity) as the light polarization changes from circular to linear (not shown here) due to the above reasons.[22] Similar transformation in the polarity is not observed in BF cell.

The resonances in ARC cell (Length ~25 mm, Temp. ~25⁰ C) have smaller amplitude compared to BF cell (Length ~50 mm, Temp ~ 48⁰ C). Thus, higher amplitude of magnetic modulation (~2 $\mu T$) is used for ARC cell as compared to ~0.14 $\mu T$ used for BF cell. This leads to increase in the width of the resonances to ~2.29 $\mu T$ as compared to 0.53 $\mu T$ for BF cell. Consequently, $\delta_{RF}$ is kept at a higher value ~ -316 kHz (-58 kHz for BF cell) for resolving the resonances in ARC cell. Similarly, amplitude of RF modulation is high (10 kHz) for ARC cell compared to 0.65 kHz for BF cell. Further, the sensitivity of the lock-in amplifier is kept at 100 µV (CPT) and 300 µV (MM) for ARC cell as compared to 300 µV (CPT) and 1 mV (MM) for BF cell. For both cell, the MM signal is stronger than the CPT signal.

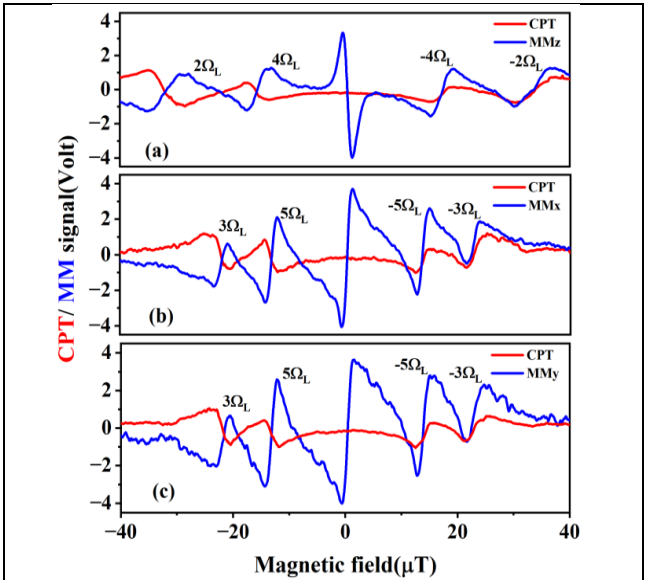

Fig-4: The simultaneous observation of CPT (red curve) and population re-distribution signal (blue curve) by scanning $B$ field in ARC cell. The orthogonal magnetic fields to the scanning direction are kept at zero. The $\delta_{RF}$ is kept fixed at ~ -316 kHz. For MM signal, a modulation of $2\mu T$ @ 39Hz is added to the scanning magnetic field. (a): The even resonances at ~ $\mp 16\mu T$ ($\pm 4\Omega_L$) and ~ $\mp 32\mu T$ ($\pm 2\Omega_L$) are obtained by scanning the $B_z$ field. (b) & (c): The odd CPT resonances at ~$\mp 13\mu T$ ($\pm 5\Omega_L$) and $\mp 22\mu T$ ($\pm 3\Omega_L$) are obtained by scanning $B_x$ and $B_y$ fields. The lock-in amplifier sensitivity is kept at 100 µV for CPT signals and 300 µV for MM signals.

The CPT resonances for the ARC cell (obtained by demodulation with respect to the modulation applied to the RF) overcome the overlap of population redistribution signal, similar to buffer gas filled cell. The resonances at even harmonics of $\Omega_L$ are observed for $B_z$ scanning where as odd harmonic resonances are observed for $B_x$ or $B_y$ scanning as shown in Fig.- 4. The $MM_x$ and $MM_y$ signals are analogous to $MM_z$ signal expect for the centrally located population redistribution signal that exhibit same slope as the quantum interference signal. It signifies enhanced transmission for $B_x$ or $B_y$ scanning in contrast to enhanced absorption for $B_z$ scanning and is consistent with Fig.-3 for BF cell.

It is interesting to note the change in slope for the positive and negative $n\Omega_L$ resonances in CPT signal (both Fig.-3 & 4). The scanning of the magnetic field from negative to positive value involves decrease in $|B|$ followed by its increase on crossing zero value. Thus the CPT resonances has opposite polarity for –ve and +ve $n\Omega_L$. Algebraically, the resonances are even function of $\nu_{RF}$ & $B$ across the centre of a resonance and line shape is independent of the ascending or descending order of any argument i.e. $f(\nu_{RF}, -B) = f(\nu_{RF}, B)$. However, the derivative profile of the signal is an odd function across the centre of a resonance and its polarity changes depending on sense of change (ascending or descending) in the argument i.e. $\left.\frac{\partial f(\nu_{RF},B)}{\partial B}\right|_{B=-B} = -\left.\frac{\partial f(\nu_{RF},B)}{\partial B}\right|_{B=+B}$. The acquired signal is the derivative of the characteristic (original) line shape due to the use of modulation-demodulation technique. Thus for scanning of the magnetic field from negative maxima to positive maxima, the slope of the signal is opposite for either side of zero magnetic field. All the resonances have same slope for RF scanning in Fig.-2(a) as the scan is unidirectional unlike magnetic field scanning. In contrast to the CPT signal, the polarity (slope) of MM signal is independent of the sign of $n\Omega_L$ resonances. It is due to an additional change in the phase of the energy level oscillation (arising from the magnetic field modulation for MM signal) for positive and negative side of the magnetic field scanning. Consequently, the polarity of the MM resonances gets preserved against sign of the $n\Omega_L$ resonance. For acquisition of CPT signal, the RF is modulated and there is no involvement of additional change in phase leading to change in the slope of the CPT resonances.

The method of scanning the magnetic field at fixed $\delta_{RF}$ is convenient for calibration of the magnetic coils and measurement of the magnetic field. The cancelation of ambient magnetic field at the Rb cell is a pre requisite for the calibration of the magnetic coils. The CPT signal vs $B$ field plot is used to estimate/cancel the component of ambient magnetic field along any axis, irrespective of presence of field along orthogonal axes. This is done by taking $B_x$, $B_y$ and $B_z$ scan of the resonances. From the symmetric presence of resonances, the ambient field along respective directions are find out and cancelled by applying bias fields. Another set of $B_x$, $B_y$ and $B_z$ scan are taken after cancelling the background magnetic field (Fig.-3 and 4) and coils are calibrated using Eq. 1-3. Since the RF detuning can be accurately known (frequency measurement) and measurements are derived from fundamental atomic constants,[23] the calibration method is highly accurate.

| Bias Field at BF cell | | | |
|---|---|---|---|
| Method | $B_x$ ($\mu T$) | $B_y$ ($\mu T$) | $B_z$ ($\mu T$) |
| RF Scanning | -0.3 | -2.3 | ~0 |
| Magnetic Scanning | -0.3 | -2.2 | ~0 |
| Zero field resonance | -0.3 | -2.2 | -0.1 |
| Calibration factor of coils in BF cell | | | |
| Method | $B_x$ (μT/V) | $B_y$ (μT/V) | $B_z$ (μT/V) |
| RF Scanning | 0.6 | 0.6 | 3.1 |
| Magnetic Scanning | 0.6 | 0.5 | 3.1 |

Tab.-1: Comparative measurement of bias field and calibration of the coils using RF and $B$ scan in BF cell. The bias field is also measured from the zero field population redistribution signal that can be observed only through $B$ field scanning.

The bias fields for compensating background magnetic field around the BF cell are compared for different methods in Tab.-1. The measurements involving conventional method of scanning RF and the modified method of scanning $B$ field are in close agreement. The centre of the zero field resonance (*MM* signal) due to population redistribution/quantum interference also provides a handle for estimation of the background magnetic fields and is compared.[24,25] However, the calibration of the coils can be performed only with resonances arising from non-degenerate states. The calibration factors calculated by both the methods are in agreement.

| Bias Field at ARC cell | | | |
|---|---|---|---|
| Method | $B_x$ ($\mu T$) | $B_y$ ($\mu T$) | $B_z$ ($\mu T$) |
| CPT resonance | 2.1 | 1.3 | 0.3 |
| Zero field resonance | 2.4 | 1.7 | 0.7 |

Tab.-2: Comparative measurement of bias field in ARC cell obtained through CPT resonance and zero field resonance, both by scanning of $B$ field.

Tab.-2 illustrate the consistent measurement of bias field in ARC cell using the CPT signal and zero field resonance of the *MM* signal, both through scanning of $B$ field. Unlike BF cell, the quantum interference resonances are prominently observed in the *MM* signal for ARC cell and can be equally used for calibration of the magnetic coils. The possibility of simultaneous acquisition of magnetic resonance at zero field (population redistribution signal) and at high field (quantum interference signal) can pave way for unexplored applications like devising unique magnetometers with large dynamic range. Further, the dependence of the slope of the CPT resonances on the polarity of the magnetic field can be gainfully utilized for vector field measurement.

## IV. Conclusions:

An adapted method of acquiring quantum interference signal by scanning magnetic field at a fixed RF detuning is discussed. Here, the quantum interference between the different Zeeman states for non-degenerate states, degenerate state, and the population redistribution signal can be acquired simultaneously. The resonances in two complementary medium, a 50 mm long $N_2$ filled cell and another 25 mm long anti-relaxation coated atomic cell are compared. The utilized method is found to be advantageous for calibration of the magnetic coils as compared to conventional method of scanning RF detuning at a fixed magnetic field. Both methods demonstrated reliable and consistent results, validating the accuracy of coil calibration that is derived using fundamental atomic constants. The adapted method can be useful for design of magnetometer with large dynamic range and having vector field measurement capability.

## V. Acknowledgment:

**Data Availability Statement:**

The data that support the findings of this study are available from the corresponding author upon reasonable request.